\documentstyle[12pt,epsf]{article}
\newcommand{\TPG}{T_{\rm PG}}

\newcommand{\Tc}{T_{\rm c}}

\newcommand{\TtG}{T_{2{\rm G}}}

\begin{document}
\begin{center}
\LARGE {\bf Pseudogap and Kinetic Pairing Under Critical Differentiation of 
Electrons in Cuprate Superconductors}\\
\vspace{5mm}
\large{Masatoshi Imada$^{1)}$ and
Shigeki Onoda}\\
\vspace{5mm}
\normalsize
{ Institute for Solid State Physics, University of Tokyo, 
5-1-5, Kashiwanoha, Kashiwa, Chiba, 277-8581, Japan}\\
{1) e-mail address, imada@issp.u-tokyo.ac.jp}
\end{center}

\baselineskip 18pt

\begin{abstract}
Superconducting mechanism of cuprates is discussed in the light of the proximity of the Mott insulator.  The proximity 
accompanied by suppression of coherence takes place in an inhomogeneous way in 
the momentum space in finite-dimensional systems.  Studies on instabilities of 
metals consisted of such differentiated electrons in the momentum space are 
reviewed from a general point of view.  A typical example of the differentiation is found in the flattening of the quasiparticle 
dispersion discovered around momenta $(\pi,0)$ and $(0,\pi)$ on 2D square 
lattices.  This flattening even controls the criticality of the metal-insulator 
transition. 
Such differentiation and suppressed coherence subsequently cause an instability 
to the superconducting state in the second order of the strong coupling 
expansion.  The d-wave pairing interaction is generated from such local but 
kinetic processes in the absence of disturbance from the coherent 
single-particle excitations.  The superconducting mechanism emerges from a direct kinetic origin which is conceptually different from the pairing mechanism mediated by bosonic excitations as in magnetic, excitonic, and BCS mechanisms. Pseudogap phenomena widely observed in the 
underdoped cuprates are then naturally understood from the mode-mode coupling of 
d-wave superconducting (dSC) fluctuations repulsively coupled with 
antiferromagnetic (AFM) ones.  When we assume the existence of a strong d-wave 
channel repulsively competing with AFM fluctuations under the formation of flat 
and damped single-particle dispersion, we reproduce basic properties of the 
pseudogap seen in the magnetic resonance, neutron scattering, angle resolved 
photoemission and tunneling measurements in the cuprates. 

\end{abstract}
\setlength{\leftmargin}{-30mm}

\section{Introduction}
Magnetism in strongly correlated electron systems has been a subject of 
extensive studies for a long time.  In many examples, the energy gain from electron 
kinetic energy term is a crucial driving force for stabilizing magnetically symmetry 
broken states.  Even in the Mott insulating state, this is equally true.  
Although the itinerancy of electrons is lost by the Coulomb interaction in the 
Mott insulator, electrons still find an optimized way to gain the energy from the kinetic energy term 
through a virtual process, where the strong coupling expansion becomes an 
adequate description under totally suppressed coherent motion of electrons.  
Anderson's mechanism of the superexchange interaction thus 
proposed~\cite{Anderson1} has long been one of the most fruitful concepts in the 
condensed matter physics.    The double exchange mechanism proposed by de 
Gennes~\cite{deGennes}, Zenner~\cite{Zener} and Anderson and 
Hasegawa~\cite{Anderson2} is another example, where a ferromagnetic (FM) metal 
is stabilized by the kinetic energy gain under a strong Hund's rule coupling 
between degenerate orbitals.

Superconductivity, on the other hand, has been analyzed mainly from a different 
view since the dramatic success of BCS theory~\cite{BCS}.  Various types of 
pairing interactions mediated by bosonic excitations have been considered to be 
responsible for the mechanisms. To say nothing of the electron-phonon 
interaction, roles of AFM and FM paramagnons and excitons have been studied in 
various contexts. In these proposals, the superconducting ground state is the 
consequence of pair formation by the mediated attraction in the presence of 
coherent and metallic quasiparticles.   

In this report, we discuss in more general perspective the importance of the 
kinetic origin for various symmetry breakings.  We particularly discuss the 
kinetic mechanism of the superconductivity under the suppressed coherence of 
quasiparticles near the Mott insulator.  
The mechanism of superconductivity we discuss in this paper is conceptually 
different from the conventional magnetic mechanism discussed in the spin 
fluctuation theories and in the $t$-$J$ model, because the origin is not in the 
mediated bosons but in the direct kinetic process as in the superexchange for 
the mechanism of antiferromagnetism.
 Although the superexchange interaction is the only process for the energy lowering by the kinetic 
energy term in the Mott insulating phase, this constraint is lost in the 
presence of doped carriers and various other processes may equally or even more 
importantly contribute to the kinetic energy gain.  The existence of various 
other processes is in fact easily confirmed by the strong coupling expansion of 
the Hubbard-type models, where in the second order, namely in the order of 
$t^2/U$, it not only generates the superexchange interaction but also other 
two-particle terms including the pair hopping processes.  Here in the strong 
coupling expansion, the onsite Coulomb repulsion $U$ is assumed to be larger 
than the kinetic transfer term $t$.  
These second-order terms usually have secondary importance in the presence of 
the kinetic energy term in the first order of $t$ in the normal metal.  However, 
as similarly to the superexchange interaction in the Mott insulating phase, they 
become the primary origin of the kinetic energy gain even in metals when the 
single-particle processes are suppressed by some reason but the amplitude of the 
second-order terms are retained.  In the study on the strong-coupling expansion 
of the d-p model, it was shown that the d-wave pair-hopping process appears in 
the third order, although the superexchange interaction appears only in the 
fourth order~\cite{Tsunetsugu99}.  In the next section we first discuss the 
origin of the unusual suppression of coherent single-particle motion in the 
proximity of the Mott insulator before considering the mechanism of 
superconductivity.

High-temperature cuprate superconductors show a variety of unusual properties in 
the normal state~\cite{RMP}.  Among others, we concentrate on two remarkable 
properties widely observed in the cuprates for the purpose of examining the 
relevance of  the kinetic mechanism as the driving mechanism of the high-Tc 
superconductivity.  One of the remarkable properties in the cuprates is the flat 
dispersion in single-particle excitations.  
Angle resolved photoemission spectra (ARPES) in Y and Bi based high-Tc  cuprates 
show an unusual dispersion which is far from weak correlation 
picture~\cite{Gofron}.  The dispersion around $(\pi,0)$ and $(0,\pi)$ is 
extremely flat beyond the expectations from usual van Hove singularities.  
The flat dispersion also shows rather strong damping.  

The other remarkable property we discuss is the pseudogap phenomenon observed in 
the underdoped region~\cite{RMP,Yasuoka}. It is observed both in spin and charge 
excitations in which gap structure emerges from a temperature $\TPG$ well above 
the superconducting transition point $\Tc$.  
The gap structure is observed in various different probes such as NMR relaxation 
time, the Knight shift, neutron scattering, tunneling, ARPES, specific heat, 
optical conductivity, and DC resistivity.  The 
ARPES~\cite{LoeserShenDessauMarshallParkFournierKapitulnik1996,photoemission} 
data have revealed that the pseudogap starts growing first in the region around 
$(\pi,0)$ and $(0,\pi)$ from $T=\TPG$ much higher than $T_c$.  
Therefore, the pseudogap appears from the momentum region of the flattened 
dispersion and it is likely that the mechanism of the pseudogap formation is 
deeply influenced from the underlying flatness.   The superconducting state 
itself also shows a dominant gap structure in this flat spots, $(\pi,0)$ and 
$(0,\pi)$, due to the $d_{x^2-y^2}$ symmetry. In fact, the pseudogap structure 
above $T_c$  appears continuously to merge into the $d_{x^2-y^2}$ gap below 
$\Tc$.  To understand the superconducting mechanism and the origin of the high 
transition temperatures, a detailed understanding of the physics taking place in 
the flat dispersion region is required.  

The emergence of the flat dispersion around $(\pi,0)$ and $(0,\pi)$ 
has also been reported in numerical simulation results rather universally in 
models for strongly correlated electrons such as the Hubbard and $t$-$J$ models 
on square lattices~\cite{Dagotto,Hanke,Assaad98}. 
As we see in the next section, this criticality is interpreted from a strong 
proximity of the Mott insulator where strong electron correlation generates 
suppressed dynamics and coherence.  

\section{Kinetic Pairing Derived from Electron Differentiation}

In a simple picture, the correlation effects emerge as the isotropic mass 
renormalization, where the Coulomb repulsion from other electrons makes the 
effective mass heavier.  This effect was first demonstrated by Brinkman and 
Rice~\cite{Brinkman} in the Gutzwiller approximation and refined in the 
dynamical mean field theory ~\cite{DMFT}.   

In the numerical results on a square lattice, as discussed above, the 
correlation effects appear in more subtle way where the electrons at different 
momenta show different renormalizations.  When the Mott insulator is approached 
and the doping concentration becomes small, the mass renormalization generally 
becomes stronger.  
However, once the renormalization effect gets relatively stronger in a part of 
the Fermi surface, it is further enhanced at that part in a selfconsistent 
fashion because the slower electrons become more and more sensitive to the 
correlation effect. This generates critical differentiation of the carriers 
depending on the portion of the Fermi surface.

On square lattices, the stronger renormalization happens around ($\pi,0$) and 
($0,\pi$). 
A part of this anisotropic correlation effect concentrating near ($\pi,0$) and 
($0,\pi$) is intuitively understood from the carrier motion under the background 
of AFM correlations.  
As we see a real space picture in Fig.1a, the carrier motion in the diagonal 
directions does not disturb the correlations due to the parallel spin alignment, 
while the motion in horizontal and vertical directions strongly disturbs the AFM 
backgrounds as expressed as the wavy bonds and the motion itself is also 
disturbed as a feedback.   
Such strong coupling of charge dynamics to spin correlations causes flattening 
and damping of electrons around ($\pi,0$) and ($0,\pi$), but not around the 
diagonal direction ($\pm {\pi}/{2},\pm {\pi}/{2}$).  
The anisotropic renormalization effect eventually may generate a singularly flat 
dispersion on particular region of the Fermi surface, which accepts more and 
more doped holes in that region due to the enhanced density of states. 

The transition to the Mott insulator is then governed by that flattened part, 
since the carriers reside predominantly in the flat region.    
The criticality of the metal-insulator transition on the square lattice is thus 
determined from the doped carriers around the flat spots, $(\pi,0)$ and 
$(0,\pi)$.  
The hyperscaling relation becomes naturally satisfied because singular points on 
the momentum space govern the transition.
In fact, the hyperscaling relations are numerically supported in various 
quantities and shows agreements with experimental indications.  For example, the 
electronic compressibility critically diverges as $\kappa \propto 1/\delta$  
with decreasing doping concentration $\delta$ while the Drude weight is 
unusually suppressed as $D\propto \delta^2$~\cite{Furukawa93,Tsune98}. 
The coherence temperature (the effective Fermi temperature) is also scaled as 
$T_F\propto \delta^2$ and indicates unusual suppression.  
In more comprehensive understanding, all the numerical data are consistent with 
the hyperscaling relations with a large dynamical exponent $z=4$ for the 
metal-insulator transition~\cite{RMP,Imada95}.  
Such large exponent opposed to the usual value $z=2$ for the transition to the 
band insulator is derived from the slower electron dynamics even at $T=0$ 
generated by the flat dispersion.

We, however, should keep in mind that the relaxation time of quasiparticles and 
the damping constant of magnetic excitations do not have criticality at the 
transition point to the Mott insulator.  
A general remark is that the relaxation time is critical only in the case of the 
Anderson localization transition and not in the case of the transition to the 
Mott insulator.  
The DC transport properties and magnetic relaxation phenomena are contaminated 
by such noncritical relaxation times $\tau$ and are influenced by the carriers 
in the other portion than the flat part because the flat part has stronger 
damping and less contributes to the DC properties.  
Large anisotropy of $\tau$ masks the real criticality and makes it difficult to 
see the real critical exponents in the $\tau$-dependent properties.  
Relevant quantities to easily estimate the criticality is the $\tau$ independent 
quantities such as the Drude weight and the compressibility.  

Near the metal-insulator transition, the critical electron differentiation and 
selective renormalization may lead to experimental observations as if internal 
degrees of freedom of the carriers such as spin and charge were separated 
because each degrees of freedom can predominantly be conveyed by carriers in 
different part of the Fermi surface.  Another possible effect of the electron 
differentiation is the appearance of several different relaxation times which 
are all originally given by a single quasiparticle relaxation in the isotropic 
Fermi liquids, but now depend on momenta of the quasiparticles.  

Another aspect one might ask in connection to the relevance of the flat part to 
the metal-insulator transition is the observed level difference between 
($\pi,0$) and ($\pm {\pi}/{2},\pm {\pi}/{2}$) in the undoped and underdoped 
cuprates~\cite{Kim}.  The level at ($\pi,0$) is substantially far from the Fermi 
level, and at a first glance, it does not have a chance to contribute to the 
metal-insulator transition.  However, it has been clarified~\cite{Tsunetsugu99} 
that this level difference is likely to be absent before the d-wave interaction 
channel starts growing where the flat part indeed governs the criticality, while 
it is developed by renormalization of the single-particle level accompanied by 
the d-wave interaction in the lower energy scale.  It is remarkable that this 
renormalization exists even in the insulator.

If the mass renormalization would happen in an isotrpic way as in the picture of 
Brinkman and Rice, the renormalization can become stronger without disturbance 
when the insulator is approached.  
However, if the singularly renormalized flat dispersion emerges critically only 
in a part near the Fermi surface but the whole band width is ratained, that 
flattened part has stronger instability due to the coupling to larger energy 
scale retained in other part of the momentum space. 
The instability can be mediated by local and incoherent carrier motion generated 
from two-particle processes derived in the strong coupling 
expansion~\cite{Tsunetsugu99}.  
The local two-particle motion is given in the order of $t^2/U$ with the bare $t$ 
while the single-particle term is renormalized to $t^*$ which can be smaller 
than $t^2/U$ at the flattened portion.
The instability of the flat dispersion was studied by taking account such local 
and incoherent terms in the Hubbard and $t$-$J$ 
models~\cite{Assaad982,Assaad983,Tsune98}.  
The inclusion of the two-particle terms drives the instability of the flat part 
to the superconducting pairing and the formation of the d-wave gap structure.    
In fact, even at half filling, the two-particle process stabilizes the d-wave 
superconducting state and reproduces the basic feature of the pseudogap 
formation~\cite{Assaad982,Assaad983} observed in the BEDT-TTF 
compounds~\cite{Kanoda}.

The paired bound particles formed from two quasiparticles at the flat spots have 
different dynamics from the original quasiparticle.  
In fact, when the paired singlet becomes the dominant carrier, the criticality 
changes from $z=4$ to $z=2$, resulting in the recovery of coherence and kinetic 
energy gain ~\cite{Tsune98}.
It generates a strong pairing interaction from the kinetic origin.  This pairing 
mechanism is a consequence of suppressed single-particle coherence and electron 
differentiation due to strong correlations.

The instability of the flat dispersion coexisting with relatively large 
incoherent process was further 
studied~\cite{Tsunetsugu99,Imada001,Imada002,Kohno00}. It has turned out that 
promotion of the above scaling behavior and the flat dispersion offers a way to 
control potential instabilities.  
Even when a flat {\it band} dispersion is designed near the Fermi level by 
controlling lattice geometry and parameters, it enlarges the critical region 
under the suppression of single-particle coherence in the
proximity of the Mott insulator mentioned above.
In designed lattices and lattices with tuned lattice parameters, it was reported 
that the superconducting instability and the formation of the spin gap have been 
dramatically enhanced~\cite{Imada001}.  

\section{Pseudogap Phenomena in Cuprates as Superconducting Fluctuations}

As is mentioned in \S1, the pseudogap in the high-Tc cuprates starts growing 
from the region of the flat dispersion.  When the single-particle coherence is 
suppressed, the system is subject to two particle instabilities.  
As clarified in \S2, the superconducting instability in fact grows.
However, the AFM and charge order correlations are in principle also expected to 
grow from other two-particle (particle-hole) processes and may compete each 
other.  In particular, the AFM long-range order is realized in the Mott 
insulator and its short-range correlation is well retained in the underdoped 
region.  Therefore, to understand how the superconducting phase appears in the 
underdoped region, at least competition of dSC and AFM correlations has to be 
treated with underlying suppressed coherence in the region of ($\pi,0$) and 
($0,\pi$).  The authors have developed a framework to treat the competition by 
employing the mode-mode coupling theory of dSC and AFM fluctuations where these 
two fluctuations are treated on an equal footing~\cite{Onoda991,Onoda992}.  

It should be noted that the strong dSC pairing interaction is resulted from a 
highly correlated effect with electron differentiation while the critical 
differentiation has not been successfully reproduced from the diagrammatic 
approach so far.  
Then, within the framework of the mode-mode coupling theory, at the starting 
point, we have assumed the existence of correlation effects leading to the 
flattened dispersion and the $d$-wave pair hopping process.  The AFM and dSC 
fluctuations are predominantly generated by the contributions from the 
quasiparticle excitations in the flattened regions ($\pi,0$) and ($0,\pi$).  
These fluctuations are treated in a set of selfconsistent equations with mode 
couplings of dSC and AFM.  From the selfconsistent solution, the pseudogap 
formation is well reproduced in a region of the parameter space.  The pseudogap 
emerges when the mode coupling between dSC and AFM is repulsive with a severe 
competition and dSC eventually dominates at low temperatures.  Such competition 
suppresses $T_c$, while above $T_c$ it produces a region where pairing 
fluctuations are large.  
This region at  $T_{PG}>T>T_c$ shows suppression of $1/T_1T$ and the pseudogap 
formation around ($\pi,0$) and ($0,\pi$) in $A(k,\omega)$.  
These reproduce the basic feature of the pseudogap phenomena experimentally 
observed in the underdoped cuprates.  
The pseudogap formation is identified as coming from the superconducting 
fluctuations.  
The momentum dependence shows that the pseudogap formation starts around 
($\pi,0$) from higher temperatures and the formation temperature becomes lower 
with increasing distance from ($\pi,0$) .
All of the above reproduce the experimental observations.  

We, however, note a richer structure of the gap formation observed in the 
transversal NMR relaxation time $T_{2G}$ and the neutron resonance peak.  
One puzzling experimental observation is that the pseudogap structure appears in 
$1/T_1T$~\cite{Yasuoka,Y124NMR,Bi2212NMR,Hggap,Hg1223NMR}, while in many cases 
$1/\TtG$, which measures ${\rm Re} \chi(Q,\omega=0)$ at $Q=(\pi,\pi)$, 
continuously increases with the decrease in temperature with no indication of 
the pseudogap.  In addition, the so called resonance peak appears in the neutron 
scattering experiments~\cite{Fong}.  
A resonance peak sharply grows at a finite frequency below $\Tc$ with some 
indications even at $T_c<T<T_{\rm PG}$.  This peak frequency $\omega^\ast$ 
decreases with lowering doping concentration implying a direct and continuous 
evolution into the AFM Bragg peak in the undoped compounds.  
The neutron and $T_{\rm 2G}$ data support the idea that the AFM fluctuations are 
suppressed around $\omega=0$ but transferred to a nonzero frequency below 
$\TPG$.  

To understand these features, a detailed consideration on damping of the 
magnetic excitations is required.  
With the increase in the pairing correlation length $\xi_d$, the pseudogap in 
$A(k,\omega)$ is developed.  
Since the damping is mainly from the overdamped Stoner excitations, the gap 
formation in $A(k,\omega)$ contributes not only to suppress growth of AFM 
correlation length $\xi_{\sigma}$ but also to reduce the magnetic damping 
because, inside the domain of the $d$-wave order, the AFM excitations are less 
scattered due to the absence of low-energy quasiparticle around ($\pi,0$).  
If the quasiparticle damping is originally large around ($\pi,0$), the damping 
$\gamma$ can be reduced dramatically upon the pseudogap formation.  
Under this circumstance, our calculated result reproduces the resonance peak and 
the increase in $1/T_{2G}$ with lowering temperature at $T>T_c$ in agreement 
with the experimental observations in YBa$_2$Cu$_3$O$_{6.63}$, 
YBa$_2$Cu$_4$O$_8$ and some other underdoped compounds~\cite{Onoda991,Onoda992}.  

A subtlety arises when the damping around ($\pi/2,\pi/2$) starts contributing.  
This is particularly true under the pseudogap formation.   If contributions from 
the 
$(\pi/2,\pi/2)$ region would be absent, the damping of the magnetic excitation 
would be strongly reduced when the pseudogap is formed around $(\pi,0)$ as we 
mentioned above.  
However, under the pseudogap formation, the damping can be determined by the 
Stoner continuum generated from the  $(\pi/2,\pi/2)$ region and can remain 
overdamped.  
This process is in fact important if the quasiparticle damping around the  
$(\pi/2,\pi/2)$ region is large as in the case of La 214 
compounds~\cite{LSCOARPES}.    
The formation of the pseudogap itself is a rather universal consequence of the 
strong coupling superconductors.  However, the actual behavior may depend on 
this damping.  
If the damping generated by the $(\pi/2,\pi/2)$ region is large, it sensitively 
destroy the resonance peak structure observed in the neutron scattering 
experimental results.  

\section{Conclusion and Discussion} \label{SECTION_summary}

Electron critical differentiation is a typical property of the proximity of the 
Mott insulator.  The flattening of the quasiparticle dispersion appears around 
momenta $(\pi,0)$ and $(0,\pi)$ on square lattices and determines the 
criticality of the metal-insulator transition with the suppressed coherence in 
that momentum region of quasiparticles. 
Such coherence suppression subsequently causes an instability to the 
superconducting state when a proper incoherent kinetic process is retained.  The d-wave 
superconducting state is stabilized from such retained microscopic process 
derived from the strong correlation expansion.  The origin of the superconductivity is ascribed to the kinetic energy gain.

By assuming the d-wave channel and the presence of strongly renormalized flat 
quasi-particle dispersion around the $(\pi,0)$ region,  we have constructed the 
mode-mode coupling theory for the AFM and $d$SC fluctuations. 
The pseudogap in the high-$\Tc$ cuprates is reproduced as the region with 
enhanced $d$SC correlations and is consistently explained from precursor effects for the superconductivity.  
The existence of the flat region plays a role to suppress the effective Fermi 
temperature $E_F$.  This suppressed $E_F$ and relatively large local pair hopping process both drive the system to the strong coupling superconductor 
thereby leading to the pseudogap formation. 
The pseudogap formation is also enhanced by the AFM fluctuations repulsively 
coupled with dSC fluctuations.

Several similar attempts have also been made to reproduce the pseudogap phenomena observed in the cuprates.  A common conclusion inferred from these 
calculations including those by the present authors is that the pseudogap is 
reproduced when the d-wave channel is explicitly 
assumed~\cite{Onoda991,Onoda992,Yamada} while crucial features such as the 
pseudogap formation around ($\pi,0$) cannot be well reproduced if one tries to 
derive the superconducting channel itself from the AFM spin 
fluctuations~\cite{Kuroda}. 
This difficulty is summarized: if one desires to stabilize a strongly fluctuating pseudogap region well above $T_c$ generated originally by the magnetic interaction,  one has to treat the strong-coupling superconductivity and hence even more strong-coupling magnetic interaction.  In such a circumstance,  it is dificult to escape from the magnetic instability before the pseudogap formation.
The difficulty is naturally interpreted from our analysis: The kinetic d-wave channel we derived in the strong coupling 
expansion is not ascribed to the magnetic origin and is not contained neither in 
the spin-fluctuation mechanism nor in the $t$-$J$ model.  This superconducting channel has a comparable amplitude to the magnetic one, namely $t^2/U$.   Furthermore the channel is enhanced by the flattened 
dispersion and the electron critical differentiation.  Such effects are 
far beyond the one-loop level of the weak coupling approach.  The change in the 
quasiparticle dispersion in the portion of the Fermi surface is not well 
reproduced for the moment and the momentum dependent selfenergy appears to be 
significantly underestimated in the existing diagrammatic evaluations.

The strong coupling Hamiltonian was considered before~\cite{Chao} and also in terms of the pairing arising from the attractive superexchange interaction~\cite{Hirsch}.  The superconductivity from direct kinetic origin was also discussed before in a specific model~\cite{Hirsch2}
as well as in the interlayer tunneling mechanism~\cite{Anderson3}, where interlayer kinetic 
energy gain is required to stabilize the superconductivity.  If we identify the 
intralayer charge incoherence observed in the normal state as the crossover 
phenomena above the unusually suppressed Fermi temperature in the proximity of 
the Mott insulator, and notice that no quantum coherence is reached even within 
the layer, the kinetic origin of the superconductivity is found solely in a 
two-dimensional plane.     

Under the suppression of coherence in the proximity of the Mott insulator, the 
second-order process in the strong-coupling expansion may also drive other type 
of ordered state in addition to the superconductivity.  In particular, because 
the charge compressibility is enhanced due to the flattened dispersion, the 
charge fluctuation may be strongly enhancced and the charge ordering and stripes 
can be triggered both by the kinetic energy gain and the intersite Coulomb 
interaction.  When the competition between the superconductivity and the charge 
ordering becomes serious, presumable repulsive mode coupling of them may become 
another origin of the pseudogap formation, although such evidence is not clear 
in the present experimental results of the cuprates. 

The presend theory of the kinetic superconductor predicts a specific form of the kinetic energy gain to be seen in the optical conductivity~\cite{Basov} and the single particle spectra~\cite{Norman}.  Qualitatively, it is expected that the in-plane kinetic energy starts gained even in the pseudogap region.  The energy gain dominantly coming from the $(\pi,0)$ and $(0,\pi)$ region of the single particles must have a significant doping dependence associated to the change in the dynamical exponent $z$ from 4 to 2.

\begin{figure}[hbt]
\begin{center}\leavevmode
\epsfxsize=12.6cm
$$\epsffile{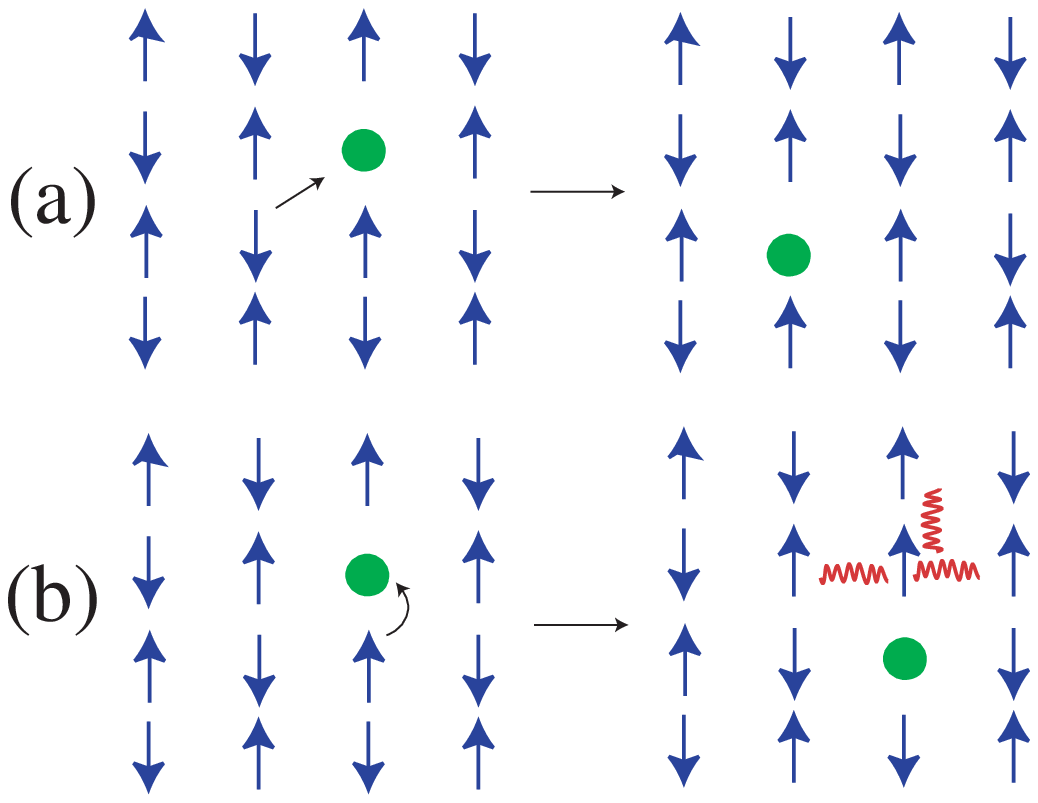}$$
\caption{Intuitive picture to understand anisotropic renormalization effects. An 
electron moving in diagonal directions under the AFM correlations are not 
severely renormalized as in (a) while they are for horizontal or vertical 
directions as in (b). In (b), frustrations are generated after the hole motion 
(denoted by circles) as shown by wavy bonds. These differences induce the 
differentiation between electrons around $(\pi,0)$ and $(\pi/2,\pi/2)$. }
\label{CRESTFIG}
\end{center}
\end{figure}
\end{document}